\documentclass[final,5p,times,twocolumn]{elsarticle}

\usepackage{geometry}
\usepackage{amssymb}
\usepackage{amsmath}
\usepackage{array}
\usepackage[hyphens]{url}
\usepackage{hyperref}
\usepackage{listings}
\usepackage{xcolor}
\usepackage{multirow}
\usepackage{tabularx}
\usepackage{booktabs}

\DeclareMathAlphabet{\mathcal}{OMS}{cmsy}{m}{n}
\SetMathAlphabet{\mathcal}{bold}{OMS}{cmsy}{b}{n}
\newcommand{\bigO}{\mathcal{O}}

\journal{}

\begin{document}

\begin{frontmatter}



\title{Multi-GPU Quantum Circuit Simulation and the Impact of Network Performance} %% Article title


\author[a]{W. Michael Brown\corref{cor1}} %% Author name
\cortext[cor1]{Corresponding Author.}
\ead{michbrown@nvidia.com}
\author[b]{Anurag Ramesh} %% Author name
\ead{rames102@purdue.edu}
\author[c,d]{Thomas Lubinski} %% Author name
\ead{tlubinski@quantumcircuits.com}
\author[a]{Thien Nguyen} %% Author name
\ead{thiennguyen@nvidia.com}
\author[b]{David E. Bernal Neira} %% Author name
\ead{dbernaln@purdue.edu}
\affiliation[a]{organization={NVIDIA},%Department and Organization
            address={2788 San Tomas Expressway},
            city={Santa Clara},
            state={CA},
            zip={95051},
            country={USA}}
  
\affiliation[b]{organization={Davidson School of Chemical Engineering, Purdue University},%Department and Organization
            address={480 Stadium Mall Dr},
            city={West Lafayette},
            state={IN},
            zip={47907},
            country={USA}}
  
\affiliation[c]{organization={QED-C Technical Advisory Committee - Standards},%Department and Organization
            address={1100 Wilson Blvd., Suite 2800},
            city={Arlington},
            state={VA},
            zip={22209},
            country={USA}}
\affiliation[d]{organization={Quantum Circuits Inc.},%Department and Organization
            address={25 Science Park},
            city={New Haven},
            state={CT},
            zip={06520},
            country={USA}}

%% Abstract
\begin{abstract}
As is intrinsic to the fundamental goal of quantum computing, classical simulation of quantum algorithms is notoriously demanding in resource requirements. Nonetheless, simulation is critical to the success of the field and a requirement for algorithm development and validation, as well as hardware design. GPU-acceleration has become standard practice for simulation, and due to the exponential scaling inherent in classical methods, multi-GPU simulation can be required to achieve representative system sizes. In this case, inter-GPU communications can bottleneck performance. In this work, we present the introduction of MPI into the QED-C Application-Oriented Benchmarks to facilitate benchmarking on HPC systems. We review the advances in interconnect technology and the APIs for multi-GPU communication. We benchmark using a variety of interconnect paths, including the recent NVIDIA Grace Blackwell NVL72 architecture that represents the first product to expand high-bandwidth GPU-specialized interconnects across multiple nodes. We show that while improvements to GPU architecture have led to speedups of over 4.5X across the last few generations of GPUs, advances in interconnect performance have had a larger impact with over 16X performance improvements in time to solution for multi-GPU simulations.
\end{abstract}

%% Keywords

\end{frontmatter}



\section{Introduction}
\label{intro}

Quantum computing has seen significant advances over the last decade, with qubit counts sufficient to run algorithms of interest, the realization of the first logical qubit for fault-tolerant computing \cite{bluvstein2024logical}, and the first claims of quantum advantage \cite{google2025observation}.
The development of new quantum algorithms capable of exploiting the advantages of quantum computing is as important as advances in the hardware itself.
While quantum computers are now more capable, classical simulation of these quantum algorithms remains critical to advancing the field.
Experiments on quantum computers are time-consuming, with many challenges resulting from the need for calibration and noise mitigation.
Fault-tolerant quantum computers might require orders of magnitude more physical qubits to achieve reliable error correction, and we expect simulation to continue to play a critical role well into the fault-tolerant era.
Of course, simulation is still used in pre-silicon assessment and post-silicon debugging for the development of new classical computers.
Those who have been involved in the supercomputing race will be very familiar with the need to validate bleeding-edge hardware and software based on expected behavior, a challenge amplified in quantum computing due to the wide variety of novel engineering designs.

Rigorous performance evaluation is critical as quantum systems transition from research prototypes to practical computational tools.
Benchmarking efforts span multiple levels of the quantum computing stack, from component-level hardware characterization~\cite{Proctor2025-cd,PracticalCharacter,cross2019validating,qiskit_measuring_quantum_volume,baldwin2022re,pelofske2022quantum} and system-level performance evaluation~\cite{proctor2022measuring,blume2020volumetric,proctor2022establishing,wack_clops_2021,wack2110quality} to compiler and toolchain measurement~\cite{kharkov2022arline,nation2025benchmarking} and algorithmic performance analysis~\cite{Supermarq,QASMbench,lubinski2023_10061574,Quark,Boixo_2018,bernal2025benchmarking}.
While each provides valuable insights within its focus area, application-oriented benchmarks are essential for evaluating end-to-end system performance on representative workloads.
In this work, we utilize the QED-C Application-Oriented Performance Benchmarks for Quantum Computing~\cite{lubinski2023_10061574,lubinski2023optimization, lubinski2024quantumalgorithmexplorationusing,chatterjee2024comprehensivecrossmodelframeworkbenchmarking,niu2025practicalframeworkassessingperformance,QEDC-App-Benchmarks}, an evolving suite of quantum algorithms and applications designed to assess performance on real-world problems.
To enable scalable benchmarking on HPC platforms, we introduce support for MPI~\cite{MPI} into the QED-C framework for distributed multi-GPU evaluation. This implementation-agnostic approach supports performance assessment across quantum programming frameworks, providing insights relevant to both near-term simulation and the development of fault-tolerant quantum systems.

GPU-acceleration can provide dramatic speedups for classical simulation and has become standard practice \cite{vallero2025state,faj2023quantum} with popular simulation frameworks such as Qiskit \cite{javadi2024quantum}, PennyLane \cite{bergholm2018pennylane}, Cirq \cite{cirq}, and CUDA-Q \cite{kim2023cuda} supporting GPUs.
At maximum GPU memory capacity, we typically measure speedups of 2-3 orders of magnitude when comparing state-of-the-art CPUs and GPUs on a single socket, due to better-optimized algorithms and higher peak hardware throughput.
While techniques for gate fusion \cite{smelyanskiy2016qhipster,zhang2021hyquas,faj2023quantum,bayraktar2023cuquantum} can result in near-peak utilization of main memory bandwidth (BW) and floating-point throughput when using a single GPU, network communications can become the bottleneck when distributing a simulation across multiple GPUs.
This is relevant as state-vector simulation, the most generally applicable simulation approach, scales exponentially in memory and computational requirements with the qubit count.

For 32-bit floating-point precision, the memory requirement for the state vector in gigabytes (GiB) is given by $2^{n-27}$ for $n$ qubits.
With this encoding, current GPUs are limited to simulating around 34 qubits (128 GiB state vector) when constrained by a single GPU's memory.
While host memory can be used to supplement storage, the need for significant HPC resources can become necessary, where multiple GPUs across multiple nodes are required either to attain the aggregate memory requirements for a given qubit count or to achieve reasonable time-to-solution for the simulation of many circuits (for example, with iterative algorithms, noisy simulation, AI-guided applications, etc.). Significant work has gone into the research and implementation of efficient multi-GPU algorithms for distributed quantum circuit simulation \cite{niwa2002general, de2007massively, zhang2015quantum, smelyanskiy2016qhipster, haner2017, jones2019quest, li2021sv, zhang2021hyquas, bayraktar2023cuquantum, jiao2023communication, xu2024atlas, westrick2024grafeyn, fu2024achieving, teranishi2025lazy, deraedt2025universalquantumsimulation50}. In these cases, the performance of data movement becomes a concern as data is transferred between distributed memory domains across different GPUs or potentially between the host memory and the GPU.

The interconnect technologies for network communications between GPUs have undergone significant advancements since the introduction of GPU acceleration to HPC.
While the first systems required a path involving the host memory, support for remote direct memory access (RDMA), first marketed by NVIDIA as ``GPUDirect'' was introduced to allow direct memory access between multiple GPUs on the same node or from a network interface controller (NIC) for internode communications \cite{gpudirect}.
To overcome the limitations of the PCI Express bus initially used for intranode multi-GPU communications, high-bandwidth GPU-specialized interconnects were first introduced under the marketing name NVLink \cite{nvlink} (NVL) (Infinity Fabric for AMD Systems).
With NVLink 4, systems combining NVIDIA Grace CPUs with GPUs feature a specialized coherent interconnect between the CPU and GPU, providing the same peak bidirectional bandwidth.
This is called NVLink-C2C.
While these specialized interconnects have been limited to communications within a node, NVIDIA has recently released Grace Blackwell NVL72 \cite{nvl72} as the first generally available system to expand this high-BW all-to-all interconnect across multiple nodes, referred to here as multi-node NVL (MNNVL).
In the case of NVL72, up to 72 GPUs across multiple nodes can be in the same NVL domain.

Table \ref{table_bw} lists the peak injection bandwidths for modern interconnect options available at writing over the last few generations of GPUs.
As one can see, there is a dramatic difference in performance spanning over an order of magnitude.
While there are many aspects of network performance that are beyond the scope of this paper, for multi-GPU distributed state-vector simulation, the bisection bandwidth of the aggregate inter-GPU interconnect will typically be the primary performance concern.

%% Use a table environment to create tables.
%% Refer following link for more details.
%% https://en.wikibooks.org/wiki/LaTeX/Tables
\begin{table}[t]%% placement specifier
%% Use tabular environment to tag the tabular data.
%% https://en.wikibooks.org/wiki/LaTeX/Tables#The_tabular_environment
\centering%% For centre alignment of tabular.
\begin{tabular}{|>{\raggedright\arraybackslash}p{0.19\linewidth}|>{\centering\arraybackslash}p{0.19\linewidth}|>{\centering\arraybackslash}p{0.08\linewidth}|>{\centering\arraybackslash}p{0.08\linewidth}|>{\centering\arraybackslash}p{0.08\linewidth}|>{\centering\arraybackslash}p{0.08\linewidth}|}%% Table column specifiers
%% Tabular cells are separated by &
\hline  Inter-connect& Peak Bidirectional BW&GPU-GPU&GPU-NIC &GPU-CPU &Inter-Node\\\hline %% A tabular row ends with \\
  PCIe 4.0& 64 GB/s& X& X& X&\\\hline
  PCIe 5.0& 128 GB/s& X& X& X&\\\hline
 PCIe 6.0& 256 GB/s& X& X& X&\\\hline
 NVLink 3& 600 GB/s& X& & &\\\hline
 NVLink 4& 900 GB/s& X& & &\textsuperscript{*} \\\hline
 NVLink C2C& \multirow{2}{*}{900 GB/s}& & & \multirow{2}{*}{X}&\\\hline
 MI350X Infinity Fabric \cite{mi300}& \multirow{3}{*}{153.6 GB/s}& \multirow{3}{*}{X}& & &\\\hline
 Slingshot 11 \cite{perlmutter}& \multirow{2}{*}{25 GB/s}& & & &\multirow{2}{*}{X}\\\hline
 Connect-X 7& \multirow{2}{*}{50 GB/s}& & & &\multirow{2}{*}{X}\\\hline
 \multirow{2}{*}{NVLink 5}& 1800 GB/s& \multirow{2}{*}{X}& & &\multirow{2}{*}{X} \\ \hline
\end{tabular}
%% Use \caption command for table caption and label.
\caption{Peak Bidirectional BW for Various Interconnects and NICs Involved in Multi-GPU Communication. \textsuperscript{*}The 4th generation of NVL was technically the first to support internode communication, however, test systems supporting internode communications were never made generally available.}\label{table_bw}
\end{table}

In this work, we introduce MPI support into the QED-C benchmark suite.
We describe some of the different APIs that can be used to exploit high-performance inter-GPU communication.
We present benchmarking results using different interconnect and API options.
In addition to demonstrating the new benchmark capabilities and showcasing the performance achievable with MNNVL, we expect the work to be helpful for understanding the requirements and options available for software developers and system architects with quantum computing workloads.

\section{Methods}
\label{methods}

\subsection{Terminology}
\label{terms}

For readers new to quantum computing, we provide brief definitions of the terminology used throughout this text.

\paragraph{Classical Bits and Qubits}
A classical bit can exist exclusively in one of two states, $0$ or $1$. 
In contrast, a quantum bit, also called \emph{qubit} can occupy a coherent superposition of these basis states. 
Formally, a qubit is represented as $\alpha \lvert 0 \rangle + \beta \lvert 1 \rangle$, where $\alpha, \beta \in \mathbb{C}$ are complex amplitudes and $|\alpha|^{2} + |\beta|^{2} = 1$. 
Upon measurement, the superposition probabilistically collapses to either $\lvert 0 \rangle$ or $\lvert 1 \rangle$, with probabilities determined by the associated amplitudes.

\paragraph{Programs and Quantum Circuits}
In classical computing, algorithms are typically expressed as sequences of deterministic instructions that act on binary variables, manipulating them using Boolean logic and gate operations.
Quantum computing adopts a related abstraction in which computation is described through quantum circuits composed of unitary operations and measurements.
Unitary operations are performed through unitary operators/gates.
For example, we say an operator $A \in\mathcal{H}$ is unitary if there exists a complex-conjugate transpose $A^{\dagger}$ such that $AA^{\dagger}= I$, where I is the identity matrix.
All unitary operations are reversible ($A^{-1} = A^{\dagger}$) and therefore there is a unique inverse operation that can reverse the system back to its original state. Table~\ref{tab:classical_quantum} summarizes the relationship between classical programs and quantum programs using a set of high-level conceptual analogies.

\begin{table}[h!]
\centering
\caption{Conceptual comparison between classical and quantum programs.}
\label{tab:classical_quantum}
\small   % Ensures caption + table font match document text
\begin{tabular}{>{\raggedright\arraybackslash}p{0.2\linewidth}p{0.2\linewidth}p{0.4\linewidth}}
\toprule
\textbf{Feature} & \textbf{Classical} & \textbf{Quantum} \\
\midrule

Basic Unit &
Bit (value 0 or 1) &
Qubit (value 0, 1, or a superposition) \\
\midrule

Computation Model &
Logic gates on bits &
Unitary quantum gates on qubits \\
\midrule

Core Principle &
Deterministic Boolean logic &
Probabilistic evolution via superposition and entanglement \\
\midrule

Processing Power &
Processes one state at a time &
Evolves many basis states in superposition \\
\bottomrule
\end{tabular}
\end{table}

\paragraph{Quantum Circuit Depth}
The \emph{depth} of a quantum circuit refers to the number of sequential layers of quantum gates when arranged so that operations acting on disjoint sets of qubits are executed in parallel. 
Circuit depth plays a role analogous to computational time in classical settings: deeper circuits enable more expressive transformations but accumulate greater noise on contemporary quantum hardware. 
Consequently, depth is a key metric in the design and feasibility of near-term quantum algorithms.

\paragraph{State-Vector Simulation}
Classical simulation of quantum algorithms commonly relies on the \emph{state-vector} representation, in which the full quantum state of an $n$-qubit system is stored as a complex vector of dimension $2^n$. 
Each quantum gate is applied through a tensor product of an appropriate unitary operator. In general, gate application typically dominates the simulation run time and due to the $\bigO(2^n)$ space complexity for the state vector, the time complexity in the general case is $\bigO(g2^n)$ with gate count $g$. The exact number of floating point operations and the associated memory access patterns depend on the type of gate, the number of qubits involved, and the index of the gate within the circuit. Therefore, the simulation times at fixed qubit and gate counts can be expected to vary depending on the specifics of the circuit. When distributing a state vector across $m$ memory domains, the number of elements exchanged between domains scales as $\bigO(2^n)$ for the simulation in the general case; some gate applications can be achieved without exchanging any elements with remote domains. As there is sufficient local memory to store the state vector for $n-log_2(m)$ qubits for $m$ memory domains, distributed algorithms can partition the circuit such that gate applications involving this subset do not require inter-GPU communications. In addition to the qubit index and number of memory domains, it is also the case that the exact number of elements exchanged will depend on the type of gate.

For a single-qubit gate, for example, gate applications involve multiplying state vector elements with a 2x2 matrix defined by the gate. Given the small dimension, it is expected that this operation would be memory-BW bound at sufficient qubit counts on most system architectures. Optimization with gate fusion can be employed to group multiple gate applications into a single operation. This does not necessarily reduce floating point operations and can actually result in some increase. However, fusion can increase linear algebra dimensions, improve memory reuse at the cache and register levels, and shift the gate applications from BW-bound to a balanced arithmetic intensity. In this case, the performance goal does not involve maximizing the fusion level, but rather tuning it for the target architecture such that the simulation is not bottlenecked by memory or compute throughput. Thus, the optimal fusion level can depend on factors such as the GPU architecture and the floating-point precision used for simulation.

For more detailed background on the general algorithms involved, we recommend references \citenum{de2007massively} and \citenum{jones2023distributed} as a start.

\paragraph{Measurement and Sampling}
Quantum measurement maps the final quantum state to a classical bitstring drawn according to the underlying probability distribution determined by the state's amplitudes. 
Because a single circuit execution yields only one sample, quantum algorithms typically require repeated execution (\emph{shots}) to estimate expectation values, objective functions, or probability distributions with sufficient statistical accuracy.

\subsection{Benchmarks}
\label{benchmarks}

In this work, we use the quantum phase estimation (QPE) \citep{shukla2025practicalquantumphaseestimation} benchmark as it represents a fundamental algorithmic component in quantum computing applications and is amenable to weak-scaling studies due to the simple scaling behavior of gate count with the number of qubits.
For a more sophisticated benchmark, we also evaluate strong scaling using a 33-qubit Transverse-field Ising model \citep{TFIM} Hamiltonian from HamLib \citep{Sawaya2024hamliblibraryof} with 2D periodic boundary conditions on a $3\times22$ triagonal lattice. As a final benchmark, we simulate random circuit samples (RCS) to better cover the general case by including irregular circuits with random connectivity.

\paragraph{Quantum Phase Estimation (QPE)}
QPE is a foundational algorithmic primitive used in quantum simulation, amplitude estimation, and eigenvalue problems.
Given a unitary operator $U$ and one of its eigenstates $\lvert \psi \rangle$, the objective is to estimate the eigenphase $\phi$ such that
\[
U \lvert \psi \rangle = e^{2\pi i \phi} \lvert \psi \rangle.
\]
The QPE circuit consists of two main components: (i) a register of qubits on which controlled powers of unitary gate $U$ are applied (e.g., $U, U^2, U^4, \ldots$), and (ii) an inverse Quantum Fourier Transform (QFT) that extracts the phase information into the computational basis.
The QED-C \citep{lubinski2024quantum} implementation samples phases that are exactly representable with $k$ qubits (i.e., $\phi = n/2^k$), enabling clean fidelity comparisons.
Increasing $k$ increases both circuit width and depth while preserving algorithmic regularity. This makes QPE a valuable reference for evaluating circuit-generation time, transpilation overhead, execution time, and fidelity as a function of system size.
Within the QED-C framework, average execution times, circuit depths, and fidelity between the measured and ideal phase distributions are recorded as key performance metrics.

\paragraph{HamLib Transverse-Field Ising Model}
 HamLib \citep{Sawaya2024hamliblibraryof} provides a standardized set of Hamiltonians, including condensed matter physics models like the Heisenberg model, the Fermi-Hubbard model, and the Transverse-field Ising model (TFIM), chemistry models of molecules such as $H_2$ and $CH_4$, and combinatorial optimization models.
 The library is designed for reproducible benchmarking of quantum algorithms and different backends.
 The size of the Hamiltonians ranges from a few qubits to large-scale instances with hundreds or even thousands of qubits.

In this work, we study the time evolution of the TFIM \citep{TFIM} implemented using Suzuki-Trotter \citep{SUZUKI1990319, Dragoi_Trotter} decomposition, producing deep circuits with nontrivial interaction structure that closely resemble realistic quantum-simulation workloads.
The TFIM is used extensively in benchmarking quantum-simulation workloads. The Hamiltonian for a system of $N$ particle spins is given by
\begin{equation}\label{eqn:tfim}
H = - J \!\!\sum_{\langle i,j \rangle} \sigma_i^z \sigma_j^z 
    - h \sum_{i=1}^{N} \sigma_i^x 
\end{equation}
where $J$ is the magnetic coupling strength, $h$ is the transverse-magnetic field amplitude, $\sigma^{x,z}$ are Pauli operators, and $\langle i,j \rangle$ denotes pairs of interacting particles.

Because the Hamiltonians directly depend on qubit count, there can be significant fluctuations in the circuit depth, and therefore, we only evaluate the 33-qubit TFIM Hamiltonian at a fixed size.
The HamLib TFIM benchmark then enables assessment of strong-scaling behavior, examining fidelity degradation, execution time growth, and resource requirements for physically meaningful many-body systems.
Together, the QPE and HamLib benchmarks provide complementary perspectives on algorithmic performance across different scaling regimes.

\paragraph{Random Circuits}
For benchmarking of random circuit samples (RCS), we use a simple Python script outside of the QED-C benchmark suite that starts with an empty circuit of $n$ qubits and loops to add $30n$ gates on randomly chosen qubits. Here, the probability of a two-qubit controlled "Z" gate is 0.25, and otherwise single-qubit gates are chosen from "H", "RX", "RY", and "RZ" as defined in the CUDA-Q documentation \cite{cudaq}. Each random circuit is simulated twice, with only the second simulation timed. This allows us to exclude just-in-time compilation of the CUDA-Q kernels, as they are cached, and compilation is not benchmarked as part of this work. A fixed random seed is used to generate consistent circuits for benchmarking in different configurations.

\bigskip

The unfused gate count for the HamLib benchmark was 2987. For QPE, the gate count ranged from 625 at 33 qubits to 856 at 39 qubits. For RCS, the gate counts were $30n$ for $n$ qubits. Default CUDA-Q settings for gate fusion were used in this work, which generally give optimal performance at higher qubit counts and resulted in the best performance for these benchmarks. For FP32, the default fusion levels are 4, 5, and 5 for Ampere, Hopper, and Blackwell, respectively. There is currently no option to determine the number of gates after fusion (and therefore, this would require instrumenting the code). To normalize ideal parallel scaling performance, we provide approximate ideal numbers that assume a uniform reduction in gate counts from fusion across different qubit/GPU counts. 

For the benchmarks here, noiseless sampling was performed with 1000 shots, and 10 circuits were timed for each data point.
The timing for the first circuit was excluded as a warm-up run.
For HamLib, 10 steps using method 3 in the QED-C benchmark implementation were enforced. For QPE, the benchmark was launched with MPI as:

\begin{lstlisting}[language=Bash,basicstyle=\ttfamily\footnotesize,frame=lines]
python -m mpi4py qpe_benchmark.py -non -s 1000
   -c 10 -n $NQUBITS -w  -a cudaq
\end{lstlisting}

and for HamLib as:

\begin{lstlisting}[language=Bash,basicstyle=\ttfamily\footnotesize,frame=lines]
python -m mpi4py
   hamlib_simulation_benchmark.py -non -s 1000
   -c 11 -n 33 -w -m 3 -steps 10
   -a cudaq
\end{lstlisting}

\subsection{Quantum Simulation Software}
\label{software}

We evaluate performance using NVIDIA CUDA-Q ~\cite{cudaq}, a framework for heterogeneous quantum-classical computing.
CUDA-Q put forward a programming model, in C++ and Python, comprising types, concepts, syntax, and
semantics, which facilitates the integration of quantum processing units (QPUs) and high-performance simulation backends into conventional computing workflows. 

In CUDA-Q, QPU code is defined as a quantum kernel, which is annotated for a custom compilation pipeline, based on MLIR, to produce an executable payload for the quantum coprocessor/simulator backend.
There are various hardware QPUs and simulators available in CUDA-Q.
Specifically, both state vector and tensor network-based simulators, based on the cuQuantum library \cite{bayraktar2023cuquantum}, are available in CUDA-Q.
Each backend, either QPU or simulator, is assigned a unique target name, allowing users to interchange backends in a hardware-agnostic manner.  

The CUDA-Q simulator backend in this work, named ``nvidia'', is a high-performance state vector simulator that supports multi-node multi-GPU distribution, as well as is capable of leveraging host memory.
As the parallel algorithms for distributed state-vector simulation are primarily implemented within the cuQuantum library, the results here are more generally applicable to other frameworks in addition to CUDA-Q (at writing, multi-GPU support in Qiskit, PennyLane, Cirq, and some others is solely supported via cuQuantum).

The software versions used were CUDA-Q 0.12 with cuQuantum 25.9.1 and QED-C Application Oriented Benchmarks Git Hash ``dd5e45a6a329''.

\subsection{APIs for multi-GPU Communication}
\label{apis}

The recommended approach for exploiting multi-GPU parallelism is with high-level libraries for inter-GPU communication.
These include GPU-aware MPI, libraries based on SHMEM \cite{bariuso1994}, or newer libraries that have been heavily optimized for AI workloads.
For NVL72, the newest releases of Unified Communication X \cite{shamis2015ucx} (UCX) underlying MPI, NVSHMEM, and NVIDIA Collective Communications Library (NCCL) have all added optimized support.
Additionally, these libraries can be used in conjunction with each other; both NVSHMEM and NCCL can be used along with MPI in the same application.
For ``ninja'' level optimizations, lower-level APIs can be used for finer control of inter-GPU communications.
For GPUDirect and NVLink, these include the CUDA interprocess communication (IPC) API and the virtual memory management (VMM) API available through the CUDA Driver.
For AMD GPUs, the direct peer GPU memory access functionality is exposed by replacing the characters ``cuda'' with ``hip'' for functions in the NVIDIA-designed IPC API.
For MNNVL, the IPC API is not supported, and the VMM API should be used.

Quantum circuit simulators have been implemented using a PGAS model supporting NVSHMEM \cite{li2021sv}, and MNNVL is designed to be efficient with a global address space across GPUs.
AI-optimized libraries, such as NCCL, can offer performance benefits due to both business prioritization and the potential freedom from optimization limitations imposed by other standards.
Both of these APIs can be particularly beneficial where fine-grain communication can be exploited for performance without involvement of the GPU host. 
Nonetheless, MPI remains most prevalent in HPC and is the high-level API used for internode communication in cuQuantum and CUDA-Q.

In GPU-aware MPI, GPU memory pointers can be passed directly to MPI functions, and the runtime will choose the optimized transport suitable for the inter-GPU interconnects on the system.
For MNNVL, this is transparent; however, for typical GPU allocations performed with \lstinline[breaklines=true, basicstyle=\ttfamily\color{black}]|cudaMalloc|, zero-copy transfers are not supported, and communications will be buffered.
This can result in a performance impact in terms of latency.
In order to achieve zero-copy transfers, developers have the option to use fabric-qualified memory allocations through the VMM API or stream-ordered allocations with a fabric-qualified memory pool.
For the former, fabric-qualified memory allocation can be achieved (using CUDA 12.5 or later) with:

\bigskip

\begin{lstlisting}[language=C++,basicstyle=\ttfamily\footnotesize,frame=lines]
cudaError_t cudaMallocFabric(void** mptr, 
                   size_t size, bool device) {
  int deviceId = 0;
  auto e = cudaGetDevice(&deviceId));

  CUmemAllocationProp prop = {};
  prop.type = CU_MEM_ALLOCATION_TYPE_PINNED;
  prop.requestedHandleTypes = 
    CU_MEM_HANDLE_TYPE_FABRIC;
  prop.location.type = 
    CU_MEM_LOCATION_TYPE_DEVICE;
  prop.location.id = deviceId;

  int cpuNumaNodeId = -1;
  if (device == false) {
    cuDeviceGetAttribute(&cpuNumaNodeId,
        U_DEVICE_ATTRIBUTE_HOST_NUMA_ID,
                               deviceId));
    prop.location.type = 
      CU_MEM_LOCATION_TYPE_HOST_NUMA;
    prop.location.id = cpuNumaNodeId;
  }
  
  size_t granularity = 0;
  e = cuMemGetAllocationGranularity(
        &granularity, &prop, 
        CU_MEM_ALLOC_GRANULARITY_MINIMUM));
  size_t alignedSize=(size+granularity-1) & 
                       ~(granularity - 1);

  CUmemGenericAllocationHandle gHandle = 0;
  e = cuMemCreate(&gHandle, alignedSize, 
                  &prop, 0);

  CUmemFabricHandle fHandle;
  e = cuMemExportToShareableHandle(&fHandle, 
       gHandle, CU_MEM_HANDLE_TYPE_FABRIC, 0);

  CUdeviceptr ptr;
  e = cuMemAddressReserve(&ptr, alignedSize,
      granularity, 0, 0);
  e = cuMemMap(ptr, alignedSize, 0, 
               gHandle, 0);

  CUmemAccessDesc accessDesc  = {};
  accessDesc.flags = 
    CU_MEM_ACCESS_FLAGS_PROT_READWRITE;
  accessDesc.location.type = 
    CU_MEM_LOCATION_TYPE_DEVICE;
  accessDesc.location.id = deviceId;
  e = cuMemSetAccess(ptr, alignedSize,
                     &accessDesc, 1);

  *mptr = (void *)ptr;
  return e;
}

cudaError_t cudaFree(void *devPtr) {
  CUmemGenericAllocationHandle gHandle = 0;
  auto e = cuMemRetainAllocationHandle(&gHandle, 
                                       devPtr);
  if (e == CUDA_SUCCESS) {
    size_t size;
    CUdeviceptr base;
    CUdeviceptr cuDevPtr = (CUdeviceptr)devPtr;
    e = cuMemGetAddressRange(&base, &size, 
                             cuDevPtr);
    e = cuMemRelease(gHandle);
    e = cuMemRelease(gHandle);
    e = cuMemUnmap(cuDevPtr, size);
    e = cuMemAddressFree(cuDevPtr, size);
  }
  return e;
}
\end{lstlisting}

\bigskip

Note that full error handling for $e$ has been omitted from these examples and that even when running on only a single node, the cuMemCreate function can fail if the system does not have properly configured IMEX channels \cite{imex}.
An alternative approach to the VMM API is to allocate MPI buffers using stream-ordered memory allocators (\lstinline[breaklines=true, basicstyle=\ttfamily\color{black}]|cudaMallocAsync()|, \lstinline[breaklines=true, basicstyle=\ttfamily\color{black}]|cudaFreeAsync()|) with a memory pool that has been created with the \lstinline[breaklines=true, basicstyle=\ttfamily\color{black}]|CU_MEM_HANDLE_TYPE_FABRIC| property (see \lstinline[breaklines=true, basicstyle=\ttfamily\color{black}]|cudaMemPoolCreate()|).
In addition to CUDA-runtime support, this also carries the typical advantages for stream-ordered allocation, including low overhead where it is difficult to avoid allocation within a loop.

As with any application using CUDA-aware MPI, CUDA-Q and cuQuantum support NVL and MNNVL so long as support in the MPI communications library is configured correctly.
However, a more optimized low-level implementation can also be exploited, configured in CUDA-Q with the \lstinline[breaklines=true, basicstyle=\ttfamily\color{black}]|CUDAQ_GPU_FABRIC| environment variable.
When enabled, MPI is still used for sharing of memory handles, some synchronization, and in the case of only intranode NVL, internode communications.
However, most of the inter-GPU communications within the NVL domain are performed using an optimized implementation with the low-level APIs.

\subsection{Multi-node Benchmarking and MPI}
\label{mpi}

MPI support in cuQuantum facilitates both decreasing time to solution for a given system size or increasing the system size beyond what will fit in the memory of a single GPU or node.
Both are achieved by distributing the state vector into smaller sub-statevectors across multiple GPUs to exploit increased memory capacity, BW, and compute throughput \cite{bayraktar2023cuquantum}.
As this distribution requires intensive communication that can occur between any pairs of GPUs (see [\citenum{jones2023distributed,de2007massively}] for details on communication patterns), we demonstrate here that network communications can quickly become a bottleneck as scaling occurs. 

With parallel state-vector simulation in cuQuantum, it is required that the number of GPUs used for a single circuit be a power of 2. cuQuantum supports supplementing GPU memory with host memory; however, it is expected that this feature will come with a performance impact.
For example, when simulating with two additional qubits using host memory, one can expect a fourfold increase in simulation time compared to using only GPU memory.
This is simply because the simulation is performed on $1/4$ of the number of GPUs.\footnote{In practice, one would only supplement with host memory when there is insufficient GPU memory available. Due to the $\bigO(2^N)$ memory requirements when adding qubits and the restriction that the number of GPUs involved in parallelization must be a power of 2, at least four times as many GPUs would be required to fit the simulation entirely in GPU memory when adding 2 qubits.}
In practice, the performance hit can be significantly worse, as all the data in host memory must be moved to the GPU for computation at a bandwidth that may be limited by host memory itself or the bus to the GPU.
While this feature is certainly valuable for fully exploiting the limits of a given system or job size, here we limited the scope to focus on GPU scalability for high performance. 

At the CUDA-Q level, MPI support for simulation is enabled through the ``mgpu'' backend option. In some cases, this support is completely transparent to the user - running with MPI will automatically distribute the state vector.
Common functions for sampling and calculation of expectation values will automatically return the correct values for each MPI task. For more sophisticated workflows, CUDA-Q includes an API for common MPI functions or users can directly make use of the MPI API through libraries within C++ or Python. 

The QED-C Application-Oriented Benchmark Suite is an open-source framework designed to facilitate community contributions and extensibility.
The modular architecture supports multiple quantum programming APIs, including Qiskit and CUDA-Q, enabling straightforward integration of new capabilities such as the MPI functionality described here.
Here, we describe the implementation details of the MPI extensions to the Python-based benchmarks, with complete source code available in the QED-C Application-Oriented Benchmarks repository~\cite{QEDC-App-Benchmarks}.

The first change is to enable transparent support for running on systems that may or may not include MPI libraries.
Here, we made the decision to enable MPI only if the mpi4py module, available for installation through PyPI, is loaded. Thus, to run with MPI, users need only to add the MPI launch wrapper and load the mpi4py module (e.g. \lstinline[language=Bash, breaklines=true, basicstyle=\ttfamily\color{black}]|mpirun -np $nGPUs python3 -m mpi4py script_name.py|).
CUDA-Q/cuQuantum will automatically assign available GPUs on a node to MPI tasks in a round-robin manner. 

We then conditionally implement some common functions, for example:

\begin{lstlisting}[language=Python,basicstyle=\ttfamily\footnotesize,frame=lines]
if "mpi4py" not in sys.modules:
    def leader():
        return True
    def barrier():
        return
    def bcast(data):
        return data
else:
    from mpi4py import MPI
    def leader():
        return rank == 0
    def barrier():
        MPI.COMM_WORLD.barrier()
        return
    def bcast(data):
        return MPI.COMM_WORLD.bcast(data, 
                                    root=0)
\end{lstlisting}

A common mistake for beginners with MPI in Python is implementing all MPI tasks competing to write to the same file.
In the QED-C benchmarks, for example, it is necessary to protect the downloading of HamLib data so that only a single MPI leader performs the download, using synchronization and broadcast so that other MPI tasks don't proceed until the data is available.
Of course, the same issue will occur with output intended for the screen. For the QED-C benchmarks, we were able to use a minimally invasive approach by redirecting the output of all but one MPI task:

\begin{lstlisting}[language=Python,basicstyle=\ttfamily\footnotesize,frame=lines]
import os, sys
rank = MPI.COMM_WORLD.Get_rank()
if rank > 0:
    f = open(os.devnull, 'w')
    sys.stdout = f
\end{lstlisting}

Care must be taken with MPI parallelization to assert that random number generation produces the same result on all MPI tasks in cases where these numbers can impact the simulated circuit and parametrization. In general, it is expected that all tasks invoke CUDA-Q calls with the same parameters, and behavior otherwise is undefined and might be difficult to debug at runtime. A final important implementation note is the use of synchronization for benchmarking. That is, we enforce a barrier before starting timing so that statistics are not influenced by implicit synchronization over variable startup costs or conditional execution.

Our inclusion of stubs for parallel operations in the QED-C benchmarks, along with use of the mpi4py Python module and cuQuantum plugins for MPI support, allows for simple interoperability with CUDA-Q across systems with or without MPI or different implementations of MPI prior to ABI standardization. Additionally, the approach should ease the transition to future support for alternative communication libraries. We have added support for a subset of the QED-C benchmarks and are expanding. Support is currently limited to CUDA-Q. With expanded support, 
another concern is interoperability across different simulation frameworks. While we expect this to be largely transparent for frameworks using cuQuantum as a backend, multi-GPU parallelization choices for other backends could force more significant changes. However, we expect these to be minor compared with the larger effort required to consistently generate quantum circuits across frameworks with significant differences in programming, compilation, and execution models.

In order to exploit low-level optimizations for NVLink in cuQuantum, the CUDA-Q environment variable \lstinline[breaklines=true, basicstyle=\ttfamily\color{black}]|CUDAQ_GPU_FABRIC| can be used to specify the network configuration. This can be set to \lstinline[breaklines=true, basicstyle=\ttfamily\color{black}]|NONE| if NVLink is not available, \lstinline[breaklines=true, basicstyle=\ttfamily\color{black}]|NVL| if intranode NVLink is available, \lstinline[breaklines=true, basicstyle=\ttfamily\color{black}]|MNNVL| for running within a single MNNVL domain, or a number that specifies the size of MNNVL domains. The latter can be used, for example, when running across multiple NVL72 racks where some communication must involve InfiniBand.

\subsection{Benchmark Systems and Methodology}
\label{systems}

We can control the interconnect paths and APIs used at runtime with environment variables.
In all cases, we set \lstinline[breaklines=true, basicstyle=\ttfamily\color{black}]|OMPI_MCA_pml="ucx"| so that OpenMPI will use UCX for communications. For systems with MNNVL \lstinline[breaklines=true, basicstyle=\ttfamily\color{black}]|UCX_CUDA_IPC_ENABLE_MNNVL = {"yes", "no"}| is used to control whether MPI internode communications exploit MNNVL or only occur over the InfiniBand network. (As of UCX 1.9, the default for this environment variable is "try" such that MNNVL will be used if available). Within cuQuantum, we can compare CUDA-aware MPI implementations to low-level API implementations using the \lstinline[breaklines=true, basicstyle=\ttfamily\color{black}]|CUDAQ_GPU_FABRIC| variable. For comparing PCIe to NVL intranode, different host systems are used. The impact of InfiniBand GPUDirect RDMA is evaluated with the \lstinline[breaklines=true, basicstyle=\ttfamily\color{black}]|UCX_IB_GPU_DIRECT_RDMA = {"y", "n"}| environment variable. For all benchmarks, MPI tasks were affinitized to $1/N$ of the logical cores on the system along with the corresponding NUMA memory; this step is critical for performance for these workloads on configurations where significant data moves through host memory interconnects to the NIC.

For CUDA-Q version 0.12 used here, we also set the \lstinline[breaklines=true, basicstyle=\ttfamily\color{black}]|CUDAQ_GLOBAL_INDEX_BITS| environment variable for multi-node simulations when \lstinline[breaklines=true, basicstyle=\ttfamily\color{black}]|CUDAQ_GPU_FABRIC| is set to \lstinline[breaklines=true, basicstyle=\ttfamily\color{black}]|NVL|. This environment variable controls the topology for hierarchical communication where some groups of MPI tasks have higher communication BW. Two values were given with the first set to $\mathrm{log_2}(N)$ and the second to $\mathrm{log_2}(P/N)$ where $N$ is the number of MPI tasks within an NVL domain and $P$ is the total number of MPI tasks. For later versions of CUDA-Q, setting this environment variable is unnecessary as the configuration is automatic based on the setting for \lstinline[breaklines=true, basicstyle=\ttfamily\color{black}]|CUDAQ_GPU_FABRIC|.

The primary system used for the studies is ``Genesis'', an in-house GB200 NVL72 rack. The system has 4 GB200 ``Blackwell'' GPUs and 2 NVIDIA Grace CPUs per node with 72 GPUs connected by all-to-all MNNVL generation 5 fabric.
In addition to MNNVL, the system is configured with one ConnectX-7 (NDR InfiniBand) 400Gb port per GPU with a fully-balanced, non-blocking fat tree topology.
For the previous generation systems, we use ``Hopper'' to refer to a single x86 server with a single 80GB HBM3 H100 GPU. ``Ampere'' refers to the generation before ``Hopper'' with eight 80GB A100 GPUs in a single x86 server connected by NVL. ``Ampere-PCI'' refers to a single x86 server with eight 80GB A100 GPUs with a PCI-express interconnect.
As a baseline multi-node system using A100 GPUs, we use the Perlmutter system at the National Energy Research Scientific Computing Center (NERSC). Perlmutter has 4 A100 GPUs per node with one Slingshot 11 ``Cassini'' NIC per GPU using a 3-hop dragonfly topology for the internode network \cite{perlmutter}.
Over 7000 GPUs are available on the entire system.
Further details of the systems are available in Table \ref{table_hw}.

%% Use a table environment to create tables.
%% Refer following link for more details.
%% https://en.wikibooks.org/wiki/LaTeX/Tables
\begin{table}[t]\tiny%% placement specifier
%% Use tabular environment to tag the tabular data.
%% https://en.wikibooks.org/wiki/LaTeX/Tables#The_tabular_environment
\centering%% For centre alignment of tabular.
\begin{tabular}{|>{\raggedright\arraybackslash}p{0.1\linewidth}|>{\centering\arraybackslash}p{0.12\linewidth}|>{\centering\arraybackslash}p{0.12\linewidth}|>{\centering\arraybackslash}p{0.12\linewidth}|>{\centering\arraybackslash}p{0.12\linewidth}|>{\centering\arraybackslash}p{0.12\linewidth}|}%% Table column specifiers
%% Tabular cells are separated by &
\hline  & Ampere PCI& Ampere &Perlmutter &Hopper &Genesis\\\hline %% A tabular row ends with \\
  GPU& A100&  A100& A100& H100&GB200\\\hline
  GPU Memory & \multirow{2}{*}{80GB HBM2e}&  \multirow{2}{*}{80GB HBM2e}& 80GB HBM2e& \multirow{2}{*}{80GB HBM3}&\multirow{2}{*}{192GB HBM3e}\\\hline
 GPUs / Node& \multirow{2}{*}{8}& \multirow{2}{*}{8}& \multirow{2}{*}{4}& \multirow{2}{*}{1}&\multirow{2}{*}{4}\\\hline
 Intranode GPU Interconnect& \multirow{2}{*}{PCIe 4.0}& \multirow{2}{*}{NVL 3}& \multirow{2}{*}{NVL 3}& \multirow{2}{*}{N/A}&\multirow{2}{*}{NVL 5}\\\hline
 \multirow{2}{*}{CPU}& 2S AMD EPYC 7742& 2S AMD EPYC 7742& 1S AMD EPYC 7763& 1S AMD EPYC 7413&2S NVIDIA Grace\\\hline
 CPU Memory& 1TB DDR4 3200& 2TB DDR4 3200& 256GB DDR4 3200& 256GB DDR4 3200&960GB LPDDR5\\\hline
 CPU-GPU Interconnect& \multirow{2}{*}{PCIe 4.0}& \multirow{2}{*}{PCIe 4.0}& \multirow{2}{*}{PCIe 4.0}& \multirow{2}{*}{PCIe 5.0}&\multirow{2}{*}{NVL 4 C2C}\\\hline
 Internode Interconnect& \multirow{2}{*}{N/A}& \multirow{2}{*}{N/A}& HPE Slingshot 11& \multirow{2}{*}{N/A}&NVL 5 and ConnectX-7\\\hline
 CUDA Driver Version& \multirow{3}{*}{570.133.20}& \multirow{3}{*}{570.133.20}&\multirow{3}{*}{550.163.01} & \multirow{3}{*}{580.95.05}&\multirow{3}{*}{580.82.07}\\ \hline
 \multirow{2}{*}{OS}& Ubuntu 22.04.5 LTS& Ubuntu 22.04.5 LTS&SUSE SLES 15 SP5 & Ubuntu 24.04.3 LTS&Ubuntu 24.04.1 LTS\\\hline
\end{tabular}
%% Use \caption command for table caption and label.
\caption{Benchmark system specifications}\label{table_hw}
\end{table}

\subsection{Profiling}
\label{profiling}

Currently, the MPI implementation in cuQuantum uses dynamic library loading with direct calls to implementation-specific MPI functions. This is intended to prevent a requirement for recompilation with specific MPI configurations on a given system. However, because it bypasses direct use of the MPI API, this approach may prevent accurate use of profiling tools that intercept MPI functions at a higher level. Therefore, we use the \lstinline[breaklines=true, basicstyle=\ttfamily\color{black}]|perf| tool for profiling with the \lstinline[breaklines=true, basicstyle=\ttfamily\color{black}]|--sort dso| option for reporting. This allows identification of the total percentage of simulation time spent in MPI. As the current implementation does not overlap communication and computation for state-vector simulation, the approach is viable without additional steps to separate exposed communication time. The \lstinline[breaklines=true, basicstyle=\ttfamily\color{black}]|--call-graph dwarf| option for profiling is used to capture non-self time from calls into other libraries. To force all communications through MPI, the low-level APIs are disabled (accomplished by setting the environment variable  
\lstinline[breaklines=true, basicstyle=\ttfamily\color{black}]|CUDAQ_GPU_FABRIC=1|). To avoid profiling time in warmup runs, the \lstinline[breaklines=true, basicstyle=\ttfamily\color{black}]|--delay| option is used with the delay time determined from the end-to-end time for simulation of a single run. The number of circuits simulated was increased to ensure at least a 60-second run time for profiling. We advise that care should be taken to evaluate any profiling overhead in the simulation time. We were unable to apply this approach to the HamLib benchmark due to excessive profiling overhead (more than 2X simulation time) associated with routines handling the JIT-compiled CUDA-Q kernels.

\section{Results}
\label{results}

\begin{figure}[t]%% placement specifier
%% Use \includegraphics command to insert graphic files. Place graphics files in 
%% working directory.
\centering%% For centre alignment of image.
\includegraphics[width=1\linewidth]{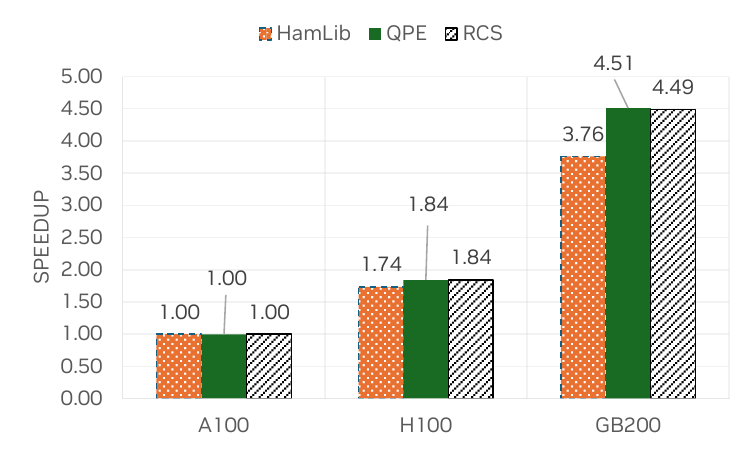}
%% Use \caption command for figure caption and label.
\caption{Single-GPU generational speedups for CUDA-Q simulation of the 33-qubit QPE, HamLib, and random circuits on GPUs from Ampere, Hopper, and Genesis systems. Absolute Ampere measurements were 3.804s, 71.589s, and 5.907s for the respective benchmarks.}\label{single_gpu}
%% https://en.wikibooks.org/wiki/LaTeX/Importing_Graphics#Importing_external_graphics
\end{figure}

\begin{figure}[t]%% placement specifier
\centering%% For centre alignment of image.
\includegraphics[width=1\linewidth]{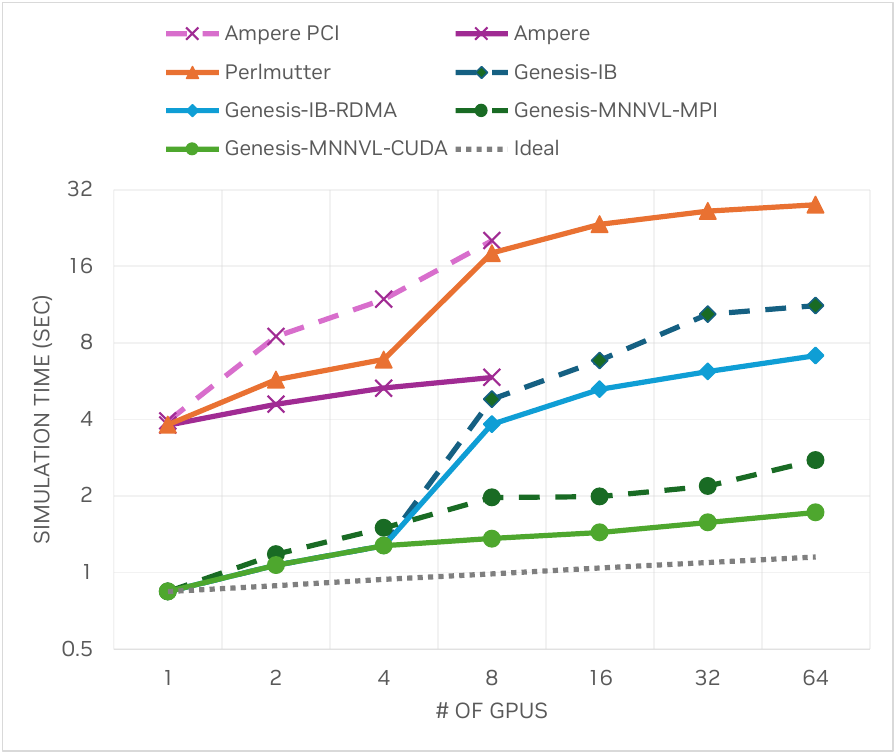}
\caption{Weak-scaling performance for the QPE benchmark on various systems. Genesis-MPI uses CUDA-aware MPI algorithms for MNNVL where Genesis-CUDA uses the low-level VMM API. Genesis-IB and Genesis-IB-RDMA disable MNNVL with the former also disabling RDMA from the NIC. The number of qubits ranges from 33 on a single GPU to 39 on 64 GPUs.}\label{qpe_weak}
\end{figure}

Single GPU generational speedups for the 33-qubit QPE, HamLib, and RCS benchmarks are shown in Figure \ref{single_gpu} with approximately 1.8X speedup going from the NVIDIA Ampere to Hopper generations and 2.2-2.4X from Hopper to Blackwell.
Multi-GPU weak scaling performance for the QPE benchmark is shown in Figure \ref{qpe_weak}, starting with 33 qubits and increasing this number by one with each doubling of GPU counts.
This represents the largest simulations fitting into GPU memory on Perlmutter.
Due to the higher memory capacity of the GB200 GPUs, we could simulate between 34 and 40 qubits on the Genesis system with better parallel efficiency.
However, this would obfuscate the comparison to previous generations.
The ideal weak scaling performance is based on simulation time for a single GB200 GPU, normalized to account for the increasing gate count for larger circuits with more qubits.

%% Use figure environment to create figures
%% Refer following link for more details.
%% https://en.wikibooks.org/wiki/LaTeX/Floats,_Figures_and_Captions

The speedup with MNNVL compared to InfiniBand on Genesis ranges from 2.8-4.1X, going from 2 to 16 nodes.
With MNNVL, there is an initial decrease in parallel efficiency to 73\% at 4 GPUs.
This is due to a significant fraction of data movement shifting from 8 TB/s HBM 3e memory to the 1.8 TB/s (bidirectional) MNNVL links.
This balances out, and parallel efficiency remains steady at 67-73\% up to 64 GPUs.
With InfiniBand, there is a significant decrease in performance going from 4 GPUs within an intranode NVL domain to 8 GPUs across nodes over an interconnect path with significantly lower BW.
As the addition of this path significantly impacts the bisection-BW, the benefit of intranode NVL can be diminished (see where ``Ampere-PCI'' meets Permutter).   

\begin{figure}[t]%% placement specifier
\centering%% For centre alignment of image.
\includegraphics[width=1\linewidth]{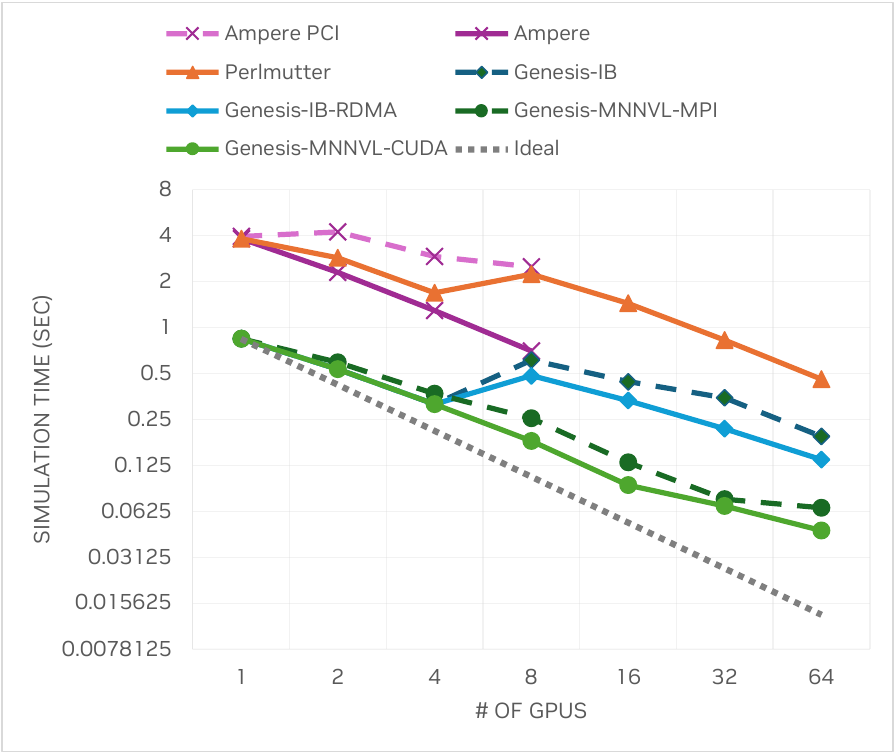}
\caption{Strong-scaling performance for the 33-Qubit QPE benchmark on various systems. Genesis-MPI uses CUDA-aware MPI algorithms for MNNVL where Genesis-CUDA uses the low-level VMM API. Genesis-IB and Genesis-IB-RDMA disable MNNVL with the former also disabling RDMA from the NIC.}\label{qpe_strong}
\end{figure}

Strong-scaling performance for the 33-qubit QPE benchmark is shown in Figure \ref{qpe_strong}.
In this case, MNNVL performance is 2.7-3.6X faster than InfiniBand on the Genesis system, and performance is monotonically increasing up to 64 GPUs.
Again, with InfiniBand, there is a significant decrease in performance at the internode threshold.
After this initial shift in data movement over InfiniBand at 8 GPUs, performance again continues to improve with the addition of more aggregate BW in both the internode and intranode paths.
However, the strong scaling efficiency is significantly impacted.
Again, we see that intranode NVL benefits can be diminished.

\begin{figure}[t]%% placement specifier
\centering%% For centre alignment of image.
\includegraphics[width=1\linewidth]{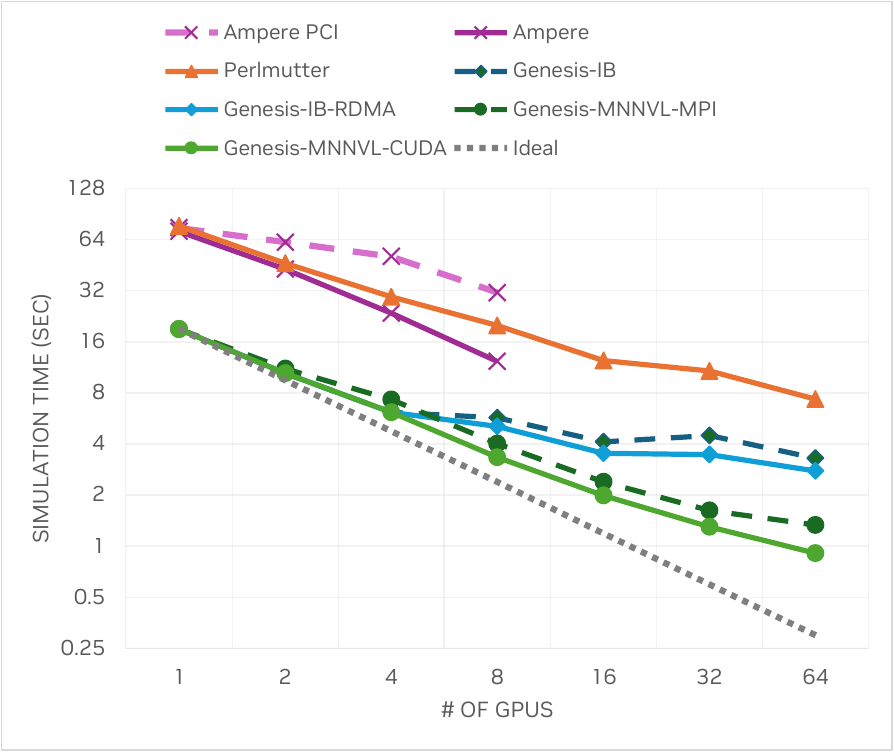}
\caption{Strong-scaling performance for the 33-Qubit HamLib benchmark on various systems. Genesis-MPI uses CUDA-aware MPI algorithms for MNNVL where Genesis-CUDA uses the low-level VMM API. Genesis-IB and Genesis-IB-RDMA disable MNNVL with the former also disabling RDMA from the NIC.}\label{hamlib}
\end{figure}

For the 33-qubit HamLib benchmark, we observe lower network sensitivity (Figure \ref{hamlib}). For a given qubit count on a specific system, lower network sensitivity for a simulation might occur for two reasons. The first is increased time in GPU kernels for gate application, where specialized kernels apply different types of gates at different fusion levels, resulting in different execution times per gate. The second is a circuit whose structure allows fewer elements to be exchanged per gate between GPUs during simulation. In this HamLib benchmark, the lower sensitivity is due to the latter. The TFIM Hamiltonian, given by equation \ref{eqn:tfim}, describes interacting particles as a linear chain with nearest-neighbor interactions. This is translated into a circuit with a ladder-like structure, allowing more gate applications to be performed without exchanging elements between MPI tasks. This results in less time spent in inter-GPU communication and thus less potential improvement from network improvements. Performance is monotonically improving in all cases, however, and MNNVL still outperforms InfiniBand by 1.5-3X on Genesis.

In contrast to the QPE and HamLib benchmarks, the random circuits exhibit irregular structure and connectivity by construction. This can be a useful test, as it is unexpected that optimizations from general distributed state vector algorithms would allow an accurate solution. When benchmarking the random samples, we observe very similar scaling behavior and parallel efficiency to those of the QPE results. MNNVL speedups over InfiniBand ranged from 2-3.71X for weak scaling and from 2.13-3.24X for the 33-qubit circuits.

Although bisection-BW is the primary concern here, we still observe a significant performance decrease when disabling RDMA between the GPUs and InfiniBand NICs (Genesis-IB-RDMA vs Genesis IB).
In the strong-scaling cases, the performance hit ranges from 13\% to 59\%; with weak-scaling, it can be as high as 68\%.
In comparing MNNVL algorithms in cuQuantum, implementations with the low-level VMM API do significantly outperform CUDA-aware MPI.
Here, the impact increases at higher GPU counts.
For HamLib, this ranged from 1.06X to 1.25X between 2 and 16 nodes. For QPE with strong scaling, the range was 1.11-1.41X, and for weak scaling, the range was 1.11-1.61X.
We expect a larger impact from algorithmic differences here compared to the benefits of zero-copy without buffering.
For single-node benchmarks, the NVL interconnect significantly outperformed PCI-express with up to 2.5X performance for the HamLib benchmark and 3.5X for QPE.

In terms of expected speedups from the network performance improvements, we can estimate an idealized speedup by profiling to determine the fraction of simulation time ($p$) spent in MPI. In this case, the ideal speedup is given by $1/(p\cdot r+(1-p))$ where $r$ is the inverse ratio of the bisection bandwidths. This estimate can be complicated by hierarchical communication when using the low-level API for intranode NVL communication. However, for QPE, we observe similar performance on the Genesis system with this disabled, so all inter-GPU communication is handled through the MPI library. This allows a simple method for profiling the InfiniBand performance. These results are shown in Table \ref{tab:profile} for the QPE benchmark on the Genesis system. In this case, we compare measured MNNVL-CUDA speedups with ideal speedups from profiling in the IB-RDMA configuration with hierarchical communications disabled (i.e., disabling the use of the low-level API by setting \lstinline[breaklines=true, basicstyle=\ttfamily\color{black}]|CUDAQ_GPU_FABRIC|). For this system, $r=1/36$, and the measured speedups are from comparison to performance using the InfiniBand internode network with only MPI (results not plotted) to account for any small performance changes from previously shown results.

\begin{table}
    \centering
    \begin{tabular}{lcccc}\toprule
         Nodes&  2&  4&  8&  16\\\midrule
         $N$-Qubit \%MPI&  68\%&  74\%&  76\%&  78\%\\
         Ideal Speedup&  2.9&  3.6&  3.8&  4.1\\
         Measured&  2.6&  3.4&  3.6&  4.0\\\midrule
         33-Qubit \%MPI&  65\%&  72\%&  70\%&  64\%\\
         Ideal Speedup&  2.7&  3.3&  3.1&  2.6\\
         Measured&  2.5&  3.7&  2.8&  2.8\\ \bottomrule
    \end{tabular}
    \caption{Ideal and measured QPE simulation speedups from profiling the percentage of time in MPI communications when using the InfiniBand network on Genesis. In this table, the low-level API is disabled for all InfiniBand measurements. Measured speedups use the CUDA low-level API for MNNVL. The top half is from weak-scaling results with 33 to 39 qubits, and the bottom half is from strong-scaling with 33 qubits.}
    \label{tab:profile}
\end{table}

Deviations from the ideal speedups (in either direction) might occur from non-uniform percentages of peak network BW, factors outside of network BW, imperfect non-network load balance, variations in hardware, and volatile state of the nodes, latencies within the communication library code, communication buffering, and the approximate nature of the profiling method. However, we observe good agreement with expectations for the algorithm as shown in Table \ref{tab:profile}. In support of the methodology, the 33-qubit and 39-qubit times with Infiniband are 0.843s and 6.965s giving an approximate parallelization cost of $6.965 - 0.843 \cdot 856/625$, or 83\% of the simulation time. While Table \ref{tab:profile} indicates 78\% of time in MPI, we expect additional parallelization overhead outside of MPI. For the InfiniBand configuration, all cases are MPI-bound with over 50\% of time spent in the library, and the speedups match expectations based on the bisection BWs. With a 36X performance improvement in BW, performance can become sensitive to implementation details in inter-GPU communication. As observed with the lower performance of the CUDA-aware MPI implementation shown in the earlier results, the low-level CUDA implementation is currently required to realize the full potential speedups. The lower percentage of time spent in MPI with strong scaling for the 33-qubit benchmark indicates lower parallel performance due to increased time spent in serial code paths and parallelization overheads that are not directly related to network performance.

\begin{figure}[t]%% placement specifier
\centering%% For centre alignment of image.
\includegraphics[width=1\linewidth]{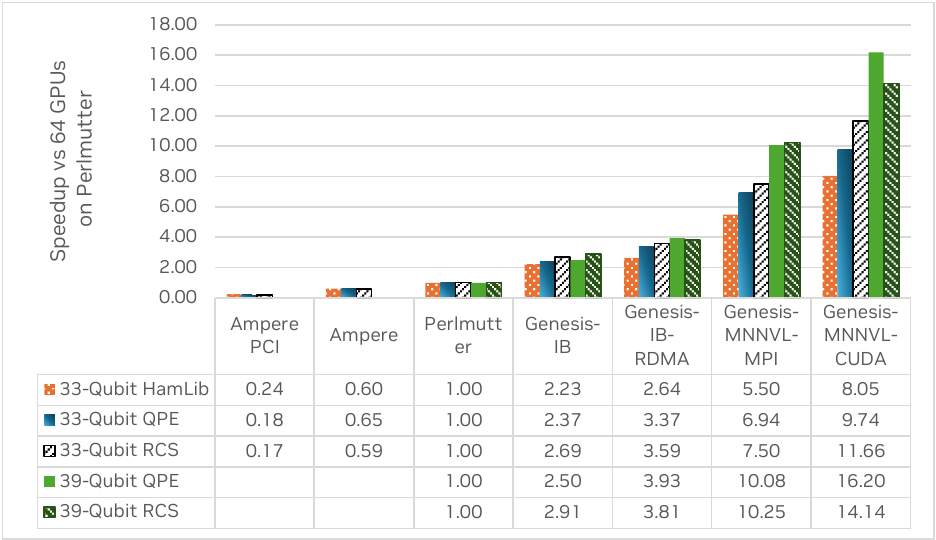}
\caption{Speedup in circuit simulation time with 64-GPU Perlmutter performance as the baseline. Absolute simulation times on Perlmutter were 7.327s, 0.458s, and 27.962s for the 33 qubit Hamlib, 33 qubit QPE, and 39 qubit QPE benchmarks. For the 33 and 39 qubit RCS benchmarks, the simulation times were 0.558s and 34.039s}\label{summary}
\end{figure}

In comparing Genesis to Perlmutter, intranode speedups with 1-4 GPUs were balanced and did not significantly deviate from those achieved with a single GPU.
Disproportionate advances in the internode interconnect with MNNVL yield larger speedups as the node count increases.
Strong-scaling performance was between 9.7X and 15X higher for multiple nodes on Genesis for QPE and 6X and 8.1X higher for HamLib.
Weak scaling performance was 13X at 36 qubits and 16.2X-16.8X at higher qubit counts. These results are somewhat higher than expectations from peak BWs, MPI profiling on Genesis (78\% at 39-qubits), and the single-GPU generational speedup (4.51X) that would give an expected approximate speedup of $(0.22 \cdot 4.51 + 0.78 \cdot 2) / (0.22 + 0.78/36) = 11X$. This indicates a lower percentage of peak interconnect BW achieved for simulations on the system, supported by a measured 77.5\% of time in MPI on Perlmutter for 39 qubits. As with the QPE benchmark, for RCS we observed higher speedups with 64 GPUs at larger qubit counts with 14.1X speedup at 39 qubits. The 33-qubit speedup is still very significant at 11.7X.
Figure \ref{summary} illustrates the speedups compared to Perlmutter, with the best simulation times for each configuration tested here.

\section{Conclusion}
\label{conclusion}

Ignoring software optimizations, over the last four years, we've seen over 4X improvement for single-GPU quantum circuit simulation with three generations of NVIDIA GPU HW.
The addition of MPI support into the QED-C benchmarks has enabled benchmarking on HPC systems using significantly higher qubit counts with better time to solution.
While the potential impact of quantum computing research is already impressive at 4X, the benchmarks show astonishing gains in multi-GPU performance, with over 16X better performance over the last three years (comparing the Grace Blackwell NVL72 system to the Perlmutter system with HPE Slingshot 11).

Although the benefits of GPU-NIC RDMA may be unclear on early systems, it is now a best practice to enable this configuration.
While this is particularly true where network latency is concerned, we still observed significant benefits with the GB200 networked through ConnectX-7 NDR InfiniBand.
In our testing, we observed expected improvements for single-node simulation using NVL as compared to PCIe for the intranode GPU interconnect.
However, this benefit can be diminished when bisection BW becomes limited at the internode interconnect. For this reason, MNNVL represents a significant advance for state-vector simulation.
As low-level algorithms in cuQuantum performed significantly better for MNNVL, users should enable this setting for best performance.

Gate fusion techniques in state-vector simulation can enable balanced arithmetic intensity with GPUs, utilizing near-peak main memory BW and floating-point throughput.
Even with MNNVL, there remains a significant gap in peak BW for main memory versus the inter-GPU interconnect.
For this reason, it is expected that a highly optimized single-GPU implementation will incur some loss in parallel efficiency; we anticipate that further optimizations for concurrent communications will be non-trivial due to increased main memory contention.
Nonetheless, there is headroom for improvement, and systems such as Genesis, with fast coherent interconnects between the host and the GPU, along with multiple internode interconnects, offer the potential for sophisticated algorithms to best exploit the system.

While our focus here has been on HW and APIs for communication, algorithmic advances will also be critical.
In this regard, sophisticated optimizations for the NVIDIA Grace-Hopper architecture were employed in the Jülich universal quantum computer simulator, demonstrating a capability for 50-qubit simulations when using the entire supercomputer \cite{deraedt2025universalquantumsimulation50}.
Of course, there are many other examples with ongoing work in-house and elsewhere \cite{jiao2023communication, rezaei2025low, teranishi2025lazy, gangapuram2024benchmarking, stenzel2024qandle, westrick2024grafeyn, adamski2023energy}.
We expect continued improvements, both in algorithms and HW, that will prove vital in advancing the field of quantum computing. 

\section{Acknowledgments}
\label{acknowledgments}

We thank Takuma Yamaguchi and Akshay Venkatesh for their expert reviews of this work. Likewise, we thank the reviewers for their time and effort to provide feedback to improve the quality of the paper.

This research used the resources of the National Energy Research
Scientific Computing Center, a DOE Office of Science User Facility
supported by the Office of Science of the U.S. Department of Energy
under Contract No. DE-AC02-05CH11231 using NERSC award
NERSC DDR-ERCAP0034389.

AR and DEBN acknowledge the support of the Davidson School of Chemical Engineering by the Center for Quantum Technologies under the Industry-University Cooperative Research Center Program at the US National Science Foundation under Grant No. 2224960.

\bibliographystyle{elsarticle-num} 
\biboptions{sort&compress}
\bibliography{references}

@article{bluvstein2024logical,
  title={Logical quantum processor based on reconfigurable atom arrays},
  author={Bluvstein, Dolev and Evered, Simon J and Geim, Alexandra A and Li, Sophie H and Zhou, Hengyun and Manovitz, Tom and Ebadi, Sepehr and Cain, Madelyn and Kalinowski, Marcin and Hangleiter, Dominik and others},
  journal={Nature},
  volume={626},
  number={7997},
  pages={58--65},
  year={2024},
  publisher={Nature Publishing Group UK London}
}

@article{google2025observation,
  title={Observation of constructive interference at the edge of quantum ergodicity},
  author={{Google Quantum AI and Collaborators}},
  journal={Nature},
  volume={646},
  number={8086},
  pages={825--830},
  year={2025},
  publisher={Nature Publishing Group UK London}
}

@article{javadi2024quantum,
  title={Quantum computing with Qiskit},
  author={Javadi-Abhari, Ali and Treinish, Matthew and Krsulich, Kevin and Wood, Christopher J and Lishman, Jake and Gacon, Julien and Martiel, Simon and Nation, Paul D and Bishop, Lev S and Cross, Andrew W and others},
  journal={arXiv preprint arXiv:2405.08810},
  year={2024}
}

@article{bergholm2018pennylane,
  title={Pennylane: Automatic differentiation of hybrid quantum-classical computations},
  author={Bergholm, Ville and Izaac, Josh and Schuld, Maria and Gogolin, Christian and Ahmed, Shahnawaz and Ajith, Vishnu and Alam, M Sohaib and Alonso-Linaje, Guillermo and AkashNarayanan, B and Asadi, Ali and others},
  journal={arXiv preprint arXiv:1811.04968},
  year={2018}
}

@book{cirq,
  url={https://zenodo.org/doi/10.5281/zenodo.4062499}, DOI={10.5281/ZENODO.4062499}, title={Cirq: Python package for writing, manipulating, and running quantum circuits on quantum computers and simulators.}, 
  publisher={Zenodo}, 
  author={Cirq Developers}, 
  year={2025}
}

@misc{gpudirect,
	author = {{\relax NVIDIA Corporation}},
	title = {{NVIDIA}’s Next Generation {CUDA} Compute Architecture: {K}epler GK110/210},
	howpublished = {\url{https://www.nvidia.com/content/dam/en-zz/Solutions/Data-Center/tesla-product-literature/NVIDIA-Kepler-GK110-GK210-Architecture-Whitepaper.pdf}},
	year = {2012},
	note = {[Accessed 31-10-2025]},
}

@inproceedings{kim2023cuda,
  title={{CUDA} {Q}uantum: The platform for integrated quantum-classical computing},
  author={Kim, Jin-Sung and McCaskey, Alex and Heim, Bettina and Modani, Manish and Stanwyck, Sam and Costa, Timothy},
  booktitle={2023 60th ACM/IEEE Design Automation Conference (DAC)},
  pages={1--4},
  year={2023},
  organization={IEEE}
}

@misc{nvlink,
	author = {{\relax NVIDIA Corporation}},
	title = {{NVIDIA} {T}esla {P100} {T}he Most Advanced Datacenter Accelerator Ever Built Featuring {P}ascal {GP100}, the World’s Fastest GPU},
	howpublished = {\url{https://images.nvidia.com/content/pdf/tesla/whitepaper/pascal-architecture-whitepaper.pdf}},
	year = {2016},
	note = {[Accessed 31-10-2025]},
}

@misc{nvl72,
	author = {{\relax NVIDIA Corporation}},
	title = {{NVIDIA} {B}lackwell Architecture Technical Brief},
	howpublished = {\url{https://resources.nvidia.com/en-us-blackwell-architecture}},
	year = {2025},
	note = {[Accessed 31-10-2025]},
}

@misc{mi300,
	author = {{\relax AMD Corporation}},
	title = {{AMD} {I}nstinct {MI300} Series Cluster Reference Architecture Guide},
	howpublished = {\url{https://www.amd.com/content/dam/amd/en/documents/instinct-tech-docs/other/instinct-mi300-series-cluster-reference-guide.pdf}},
	year = {2025},
	note = {[Accessed 31-10-2025]},
}

@misc{perlmutter,
	author = {{\relax National Energy Research Scientific Computing Center}},
	title = {Perlmutter Architecture},
	howpublished = {\url{https://docs.nersc.gov/systems/perlmutter/architecture/}},
	year = {2023},
	note = {[Accessed 31-10-2025]},
}

@misc{imex,
	author = {{\relax NVIDIA Corporation}},
	title = {{N}{V}{I}{D}{I}{A} {I}{M}{E}{X} {S}ervice for {N}{V}{L}ink {N}etworks},
	howpublished = {\url{https://docs.nvidia.com/multi-node-nvlink-systems/imex-guide/imexchannels.html}},
	year = {2025},
	note = {[Accessed 31-10-2025]},
}

@inproceedings{bayraktar2023cuquantum,
  title={cuquantum sdk: A high-performance library for accelerating quantum science},
  author={Bayraktar, Harun and Charara, Ali and Clark, David and Cohen, Saul and Costa, Timothy and Fang, Yao-Lung L and Gao, Yang and Guan, Jack and Gunnels, John and Haidar, Azzam and others},
  booktitle={2023 IEEE International Conference on Quantum Computing and Engineering (QCE)},
  volume={1},
  pages={1050--1061},
  year={2023},
  organization={IEEE}
}

@inproceedings{shamis2015ucx,
  title={{UCX}: an open source framework for {HPC} network {APIs} and beyond},
  author={Shamis, Pavel and Venkata, Manjunath Gorentla and Lopez, M Graham and Baker, Matthew B and Hernandez, Oscar and Itigin, Yossi and Dubman, Mike and Shainer, Gilad and Graham, Richard L and Liss, Liran and others},
  booktitle={2015 IEEE 23rd Annual Symposium on High-Performance Interconnects},
  pages={40--43},
  year={2015},
  organization={IEEE}
}

@book{bariuso1994,
  title     = "{SHMEM's} User's Guide",
  author    = "Bariuso, R. and Knies, A.",
  year      = 1994,
  publisher = "Cray Research, Inc.",
  address   = "Minneapolis, Minnesota"
}

@article{vallero2025state,
  title={State of practice: evaluating gpu performance of state vector and tensor network methods},
  author={Vallero, Marzio and Rech, Paolo and Vella, Flavio},
  journal={Future Generation Computer Systems},
  pages={107927},
  year={2025},
  publisher={Elsevier}
}

@misc{deraedt2025universalquantumsimulation50,
      title={Universal Quantum Simulation of 50 Qubits on Europe`s First Exascale Supercomputer Harnessing Its Heterogeneous {CPU-GPU} Architecture}, 
      author={Hans De Raedt and Jiri Kraus and Andreas Herten and Vrinda Mehta and Mathis Bode and Markus Hrywniak and Kristel Michielsen and Thomas Lippert},
      year={2025},
      eprint={2511.03359},
      archivePrefix={arXiv},
      primaryClass={quant-ph},
      url={https://arxiv.org/abs/2511.03359}, 
}

@article{smelyanskiy2016qhipster,
  title={q{H}iPSTER: The quantum high performance software testing environment},
  author={Smelyanskiy, Mikhail and Sawaya, Nicolas PD and Aspuru-Guzik, Al{\'a}n},
  journal={arXiv preprint arXiv:1601.07195},
  year={2016}
}

@inproceedings{faj2023quantum,
  title={Quantum computer simulations at warp speed: Assessing the impact of {GPU} acceleration: A case study with {IBM} {Qiskit} {AER}, {NVIDIA} {T}hrust \& {cuQuantum}},
  author={Faj, Jennifer and Peng, Ivy and Wahlgren, Jacob and Markidis, Stefano},
  booktitle={2023 IEEE 19th International Conference on e-Science (e-Science)},
  pages={1--10},
  year={2023},
  organization={IEEE}
}

@inproceedings{zhang2021hyquas,
  title={{H}y{Q}uas: hybrid partitioner based quantum circuit simulation system on {GPU}},
  author={Zhang, Chen and Song, Zeyu and Wang, Haojie and Rong, Kaiyuan and Zhai, Jidong},
  booktitle={Proceedings of the 35th ACM International Conference on Supercomputing},
  pages={443--454},
  year={2021}
}

@article{jones2023distributed,
  title={Distributed simulation of statevectors and density matrices},
  author={Jones, Tyson and Koczor, B{\'a}lint and Benjamin, Simon C},
  journal={arXiv preprint arXiv:2311.01512},
  year={2023}
}

@inproceedings{jiao2023communication,
  title={Communication optimizations for state-vector quantum simulator on {CPU+} {GPU} clusters},
  author={Jiao, Chenyang and Zhang, Weihua and Shen, Li},
  booktitle={Proceedings of the 52nd International Conference on Parallel Processing},
  pages={203--212},
  year={2023}
}

@article{rezaei2025low,
  title={Low-Level and {NUMA}-Aware Optimization for High-Performance Quantum Simulation},
  author={Rezaei, Ali and Jaulmes, Luc and Bahna, Maria and Brown, Oliver Thomson and Barbalace, Antonio},
  journal={arXiv preprint arXiv:2506.09198},
  year={2025}
}

@article{teranishi2025lazy,
  title={Lazy Qubit Reordering for Accelerating Parallel State-Vector-based Quantum Circuit Simulation},
  author={Teranishi, Yusuke and Hiraoka, Shoma and Mizukami, Wataru and Okita, Masao and Ino, Fumihiko},
  journal={ACM Transactions on Quantum Computing},
  volume={6},
  number={4},
  pages={1--33},
  year={2025},
  publisher={ACM New York, NY}
}

@article{gangapuram2024benchmarking,
  title={Benchmarking quantum computer simulation software packages: State vector simulators},
  author={Gangapuram, Amit Jamadagni and L{\"a}uchli, Andreas and Hempel, Cornelius},
  journal={SciPost Physics Core},
  volume={7},
  number={4},
  pages={075},
  year={2024}
}

@article{stenzel2024qandle,
  title={Qandle: Accelerating State Vector Simulation Using Gate-Matrix Caching and Circuit Splitting},
  author={Stenzel, Gerhard and Zielinski, Sebastian and K{\"o}lle, Michael and Altmann, Philipp and N{\"u}{\ss}lein, Jonas and Gabor, Thomas},
  journal={arXiv preprint arXiv:2404.09213},
  year={2024}
}

@inproceedings{westrick2024grafeyn,
  title={Grafeyn: Efficient parallel sparse simulation of quantum circuits},
  author={Westrick, Sam and Liu, Pengyu and Kang, Byeongjee and McDonald, Colin and Rainey, Mike and Xu, Mingkuan and Arora, Jatin and Ding, Yongshan and Acar, Umut A},
  booktitle={2024 IEEE International Conference on Quantum Computing and Engineering (QCE)},
  volume={1},
  pages={1132--1142},
  year={2024},
  organization={IEEE}
}

@inproceedings{adamski2023energy,
  title={Energy efficiency of quantum statevector simulation at scale},
  author={Adamski, Jakub and Richings, James P and Brown, Oliver Thomson},
  booktitle={Proceedings of the SC'23 Workshops of the International Conference on High Performance Computing, Network, Storage, and Analysis},
  pages={1871--1875},
  year={2023}
}

@inproceedings{li2021sv,
  title={{SV-Sim:} scalable {PGAS}-based state vector simulation of quantum circuits},
  author={Li, Ang and Fang, Bo and Granade, Christopher and Prawiroatmodjo, Guen and Heim, Bettina and Roetteler, Martin and Krishnamoorthy, Sriram},
  booktitle={Proceedings of the International Conference for High Performance Computing, Networking, Storage and Analysis},
  pages={1--14},
  year={2021}
}

@misc{cudaq,
  title = {{CUDA-Q}},
  author = {{The CUDA-Q development team}},
  year = {2025},
  url = {https://github.com/NVIDIA/cuda-quantum},
  howpublished = {\url{https://github.com/NVIDIA/cuda-quantum}},
  note = {Available at \url{https://github.com/NVIDIA/cuda-quantum}}
}

@article{Sawaya2024hamliblibraryof,
  doi = {10.22331/q-2024-12-11-1559},
  url = {https://doi.org/10.22331/q-2024-12-11-1559},
  title = {Ham{L}ib: {A} library of {H}amiltonians for benchmarking quantum algorithms and hardware},
  author = {Sawaya, Nicolas PD and Marti-Dafcik, Daniel and Ho, Yang and Tabor, Daniel P and Neira, David E Bernal and Magann, Alicia B and Premaratne, Shavindra and Dubey, Pradeep and Matsuura, Anne and Bishop, Nathan and Jong, Wibe A de and Benjamin, Simon and Parekh, Ojas and Tubman, Norm and Klymko, Katherine and Camps, Daan},
  journal = {{Quantum}},
  issn = {2521-327X},
  publisher = {{Verein zur F{\"{o}}rderung des Open Access Publizierens in den Quantenwissenschaften}},
  volume = {8},
  pages = {1559},
  month = dec,
  year = {2024}
}

@article{TFIM,
  title = {Ising Model with a Transverse Field},
  author = {Elliott, R. J. and Pfeuty, P. and Wood, C.},
  journal = {Phys. Rev. Lett.},
  volume = {25},
  issue = {7},
  pages = {443--446},
  numpages = {0},
  year = {1970},
  month = {Aug},
  publisher = {American Physical Society},
  doi = {10.1103/PhysRevLett.25.443},
  url = {https://link.aps.org/doi/10.1103/PhysRevLett.25.443}
}

@article{SUZUKI1990319,
title = {Fractal decomposition of exponential operators with applications to many-body theories and Monte Carlo simulations},
journal = {Physics Letters A},
volume = {146},
number = {6},
pages = {319-323},
year = {1990},
issn = {0375-9601},
doi = {https://doi.org/10.1016/0375-9601(90)90962-N},
url = {https://www.sciencedirect.com/science/article/pii/037596019090962N},
author = {Masuo Suzuki}
}

@article{Dragoi_Trotter,
  author    = {Sabina Dragoi},
  title     = {Analysis of the Trotter Method for Hamiltonian Simulation},
  journal   = {arXiv preprint},
  year      = {2022}
}

@misc{shukla2025practicalquantumphaseestimation,
      title={Towards Practical Quantum Phase Estimation: A Modular, Scalable, and Adaptive Approach}, 
      author={Alok Shukla and Prakash Vedula},
      year={2025},
      eprint={2507.22460},
      archivePrefix={arXiv},
      primaryClass={quant-ph},
      url={https://arxiv.org/abs/2507.22460}, 
}

@misc{wack_clops_2021,
  doi = {10.48550/ARXIV.2110.14108},
  url = {https://arxiv.org/abs/2110.14108},
  author = {Wack, Andrew and Paik, Hanhee and Javadi-Abhari, Ali and Jurcevic, Petar and Faro, Ismael and Gambetta, Jay M. and Johnson, Blake R.},
  title = {Quality, Speed, and Scale: three key attributes to measure the performance of near-term quantum computers},
  publisher = {arXiv},
  year = {2021},
  copyright = {Creative Commons Attribution 4.0 International}
}

@article{Boixo_2018,
   title={Characterizing quantum supremacy in near-term devices},
   volume={14},
   ISSN={1745-2481},
   url={http://dx.doi.org/10.1038/s41567-018-0124-x},
   DOI={10.1038/s41567-018-0124-x},
   number={6},
   journal={Nature Physics},
   publisher={Springer Science and Business Media LLC},
   author={Boixo, Sergio and Isakov, Sergei V. and Smelyanskiy, Vadim N. and Babbush, Ryan and Ding, Nan and Jiang, Zhang and Bremner, Michael J. and Martinis, John M. and Neven, Hartmut},
   year={2018},
   month={Apr},
   pages={595–600}
}

@article{lubinski2024quantum,
  title={Quantum Algorithm Exploration using Application-Oriented Performance Benchmarks},
  author={Lubinski, Thomas and Goings, Joshua J and Mayer, Karl and Johri, Sonika and Reddy, Nithin and Mehta, Aman and Bhatia, Niranjan and Rappaport, Sonny and Mills, Daniel and Baldwin, Charles H and others},
  journal={arXiv preprint arXiv:2402.08985},
  year={2024}
}

@ARTICLE{lubinski2023_10061574,
  author={Lubinski, Thomas and Johri, Sonika and Varosy, Paul and Coleman, Jeremiah and Zhao, Luning and Necaise, Jason and Baldwin, Charles H. and Mayer, Karl and Proctor, Timothy},
  journal={IEEE Transactions on Quantum Engineering}, 
  title={Application-Oriented Performance Benchmarks for Quantum Computing}, 
  year={2023},
  volume={4},
  number={},
  pages={1-32},
  doi={10.1109/TQE.2023.3253761}
}

@misc{lubinski2023optimization,
    title={Optimization Applications as Quantum Performance Benchmarks}, 
    author={Thomas Lubinski and Carleton Coffrin and Catherine McGeoch and Pratik Sathe and Joshua Apanavicius and David E. Bernal Neira},
    url={https://arxiv.org/abs/2302.02278},
    year={2023},
    eprint={2302.02278},
    archivePrefix={arXiv},
    primaryClass={quant-ph},
    doi={10.48550/arXiv.2302.02278}
}

@misc{lubinski2024quantumalgorithmexplorationusing,
      title={Quantum Algorithm Exploration using Application-Oriented Performance Benchmarks}, 
      author={Thomas Lubinski and Joshua J. Goings and Karl Mayer and Sonika Johri and Nithin Reddy and Aman Mehta and Niranjan Bhatia and Sonny Rappaport and Daniel Mills and Charles H. Baldwin and Luning Zhao and Aaron Barbosa and Smarak Maity and Pranav S. Mundada},
      year={2024},
      eprint={2402.08985},
      archivePrefix={arXiv},
      primaryClass={quant-ph},
      url={https://arxiv.org/abs/2402.08985}, 
}

@misc{chatterjee2024comprehensivecrossmodelframeworkbenchmarking,
      title={A Comprehensive Cross-Model Framework for Benchmarking the Performance of Quantum Hamiltonian Simulations}, 
      author={Avimita Chatterjee and Sonny Rappaport and Anish Giri and Sonika Johri and Timothy Proctor and David E. Bernal Neira and Pratik Sathe and Thomas Lubinski},
      year={2024},
      eprint={2409.06919},
      archivePrefix={arXiv},
      primaryClass={quant-ph},
      url={https://arxiv.org/abs/2409.06919}, 
}

@misc{niu2025practicalframeworkassessingperformance,
      title={A Practical Framework for Assessing the Performance of Observable Estimation in Quantum Simulation}, 
      author={Siyuan Niu and Efekan Kökcü and Sonika Johri and Anurag Ramesh and Avimita Chatterjee and David E. {Bernal Neira} and Daan Camps and Thomas Lubinski},
      year={2025},
      eprint={2504.09813},
      archivePrefix={arXiv},
      primaryClass={quant-ph},
      url={https://arxiv.org/abs/2504.09813}, 
}

@article{cross2019validating,
  title={Validating quantum computers using randomized model circuits},
  author={Cross, Andrew W and Bishop, Lev S and Sheldon, Sarah and Nation, Paul D and Gambetta, Jay M},
  journal={Physical Review A},
  volume={100},
  number={3},
  pages={032328},
  year={2019},
  publisher={APS}
}

@misc{qiskit_measuring_quantum_volume,
    title={Measuring quantum volume}, url={https://qiskit.org/textbook/ch-quantum-hardware/measuring-quantum-volume.html},
    journal={Qiskit}, 
    publisher={Data 100 at UC Berkeley}, 
    author={Team, The Qiskit}, 
    year={2021}, 
    month={Aug}
}

@article{baldwin2022re,
  title={Re-examining the quantum volume test: Ideal distributions, compiler optimizations, confidence intervals, and scalable resource estimations},
  author={Baldwin, Charles H and Mayer, Karl and Brown, Natalie C and Ryan-Anderson, Ciar{\'a}n and Hayes, David},
  journal={Quantum},
  volume={6},
  pages={707},
  year={2022},
  publisher={Verein zur F{\"o}rderung des Open Access Publizierens in den Quantenwissenschaften}
}

@article{pelofske2022quantum,
  title={Quantum volume in practice: What users can expect from nisq devices},
  author={Pelofske, Elijah and B{\"a}rtschi, Andreas and Eidenbenz, Stephan},
  journal={IEEE Transactions on Quantum Engineering},
  volume={3},
  pages={1--19},
  year={2022},
  publisher={IEEE}
}

@article{proctor2022measuring,
  title={Measuring the capabilities of quantum computers},
  author={Proctor, Timothy and Rudinger, Kenneth and Young, Kevin and Nielsen, Erik and Blume-Kohout, Robin},
  journal={Nature Physics},
  volume={18},
  number={1},
  pages={75--79},
  year={2022},
  publisher={Nature Publishing Group UK London}
}

@article{blume2020volumetric,
  title={A volumetric framework for quantum computer benchmarks},
  author={Blume-Kohout, Robin and Young, Kevin C},
  journal={Quantum},
  volume={4},
  pages={362},
  year={2020},
  publisher={Verein zur F{\"o}rderung des Open Access Publizierens in den Quantenwissenschaften}
}

@article{proctor2022establishing,
  title={Establishing trust in quantum computations},
  author={Proctor, Timothy and Seritan, Stefan and Nielsen, Erik and Rudinger, Kenneth and Young, Kevin and Blume-Kohout, Robin and Sarovar, Mohan},
  journal={arXiv preprint arXiv:2204.07568},
  year={2022}
}

@article{wack2110quality,
  title={Quality, speed, and scale: three key attributes to measure the performance of near-term quantum computers (2021)},
  author={Wack, Andrew and Paik, Hanhee and Javadi-Abhari, Ali and Jurcevic, Petar and Faro, Ismael and Gambetta, Jay M and Johnson, Blake R},
  journal={arXiv preprint arXiv:2110.14108},
  year={2021}
}

@article{PracticalCharacter,
  title = {Practical Introduction to Benchmarking and Characterization of Quantum Computers},
  author = {Hashim, Akel and Nguyen, Long B. and Goss, Noah and Marinelli, Brian and Naik, Ravi K. and Chistolini, Trevor and Hines, Jordan and Marceaux, J.P. and Kim, Yosep and Gokhale, Pranav and Tomesh, Teague and Chen, Senrui and Jiang, Liang and Ferracin, Samuele and Rudinger, Kenneth and Proctor, Timothy and Young, Kevin C. and Siddiqi, Irfan and Blume-Kohout, Robin},
  journal = {PRX Quantum},
  volume = {6},
  issue = {3},
  pages = {030202},
  numpages = {132},
  year = {2025},
  month = {Aug},
  publisher = {American Physical Society},
  doi = {10.1103/PRXQuantum.6.030202},
  url = {https://link.aps.org/doi/10.1103/PRXQuantum.6.030202}
}

@ARTICLE{Proctor2025-cd,
  title     = "Benchmarking quantum computers",
  author    = "Proctor, Timothy and Young, Kevin and Baczewski, Andrew D and
               Blume-Kohout, Robin",
  journal   = "Nat. Rev. Phys.",
  publisher = "Springer Science and Business Media LLC",
  volume    =  7,
  number    =  2,
  pages     = "105--118",
  month     =  jan,
  year      =  2025,
  url       = "https://www.nature.com/articles/s42254-024-00796-z",
  doi       = "10.1038/s42254-024-00796-z",
  issn      = "2522-5820,2522-5820",
}

@INPROCEEDINGS{Supermarq,
  author={Tomesh, Teague and Gokhale, Pranav and Omole, Victory and Ravi, Gokul Subramanian and Smith, Kaitlin N. and Viszlai, Joshua and Wu, Xin-Chuan and Hardavellas, Nikos and Martonosi, Margaret R. and Chong, Frederic T.},
  booktitle={2022 IEEE International Symposium on High-Performance Computer Architecture (HPCA)}, 
  title={SupermarQ: A Scalable Quantum Benchmark Suite}, 
  year={2022},
  volume={},
  number={},
  pages={587-603},
  doi={10.1109/HPCA53966.2022.00050}}

@article{QASMbench,
author = {Li, Ang and Stein, Samuel and Krishnamoorthy, Sriram and Ang, James},
title = {QASMBench: A Low-Level Quantum Benchmark Suite for NISQ Evaluation and Simulation},
year = {2023},
issue_date = {June 2023},
publisher = {Association for Computing Machinery},
address = {New York, NY, USA},
volume = {4},
number = {2},
url = {https://doi.org/10.1145/3550488},
doi = {10.1145/3550488},
journal = {ACM Transactions on Quantum Computing},
month = feb,
articleno = {10},
numpages = {26}
}

@article{nation2025benchmarking,
  title={Benchmarking the performance of quantum computing software for quantum circuit creation, manipulation and compilation},
  author={Nation, Paul D and Saki, Abdullah Ash and Brandhofer, Sebastian and Bello, Luciano and Garion, Shelly and Treinish, Matthew and Javadi-Abhari, Ali},
  journal={Nature Computational Science},
  pages={1--9},
  year={2025},
  publisher={Nature Publishing Group US New York}
}

@INPROCEEDINGS{Quark,
  author={Finžgar, Jernej Rudi and Ross, Philipp and Hölscher, Leonhard and Klepsch, Johannes and Luckow, Andre},
  booktitle={2022 IEEE International Conference on Quantum Computing and Engineering (QCE)}, 
  title={QUARK: A Framework for Quantum Computing Application Benchmarking}, 
  year={2022},
  volume={},
  number={},
  pages={226-237},
  doi={10.1109/QCE53715.2022.00042}}

@article{kharkov2022arline,
  title={Arline benchmarks: Automated benchmarking platform for quantum compilers},
  author={Kharkov, Y and Mikhantiev, E and Kotelnikov, A},
  journal={arXiv preprint arXiv:2202.14025},
  year={2022}
}

@misc{QEDC-App-Benchmarks,
    author = {{QED-C}},
    title = {{Application-Oriented Benchmarks for Quantum Computing}},
    year = {2019},
    url = {https://github.com/SRI-International/QC-App-Oriented-Benchmarks}
}

@misc{MPI,
    author = {{Multiple}},
    title = {Message Passing Interface - High Performance Computing},
    year = {2022},
    url = {https://hpc.nmsu.edu/discovery/mpi/introduction/}
}

@article{bernal2025benchmarking,
  title={{Benchmarking the operation of quantum heuristics and Ising machines: scoring parameter setting strategies on optimization applications}},
  author={Bernal Neira, David E and Brown, Robin and Sathe, Pratik and Wudarski, Filip and Pavone, Marco and Rieffel, Eleanor and Venturelli, Davide},
  journal={Quantum Machine Intelligence},
  volume={7},
  number={2},
  pages={1--12},
  year={2025},
  publisher={Springer}
}

@article{de2007massively,
  title={Massively parallel quantum computer simulator},
  author={De Raedt, Koen and Michielsen, Kristel and De Raedt, Hans and Trieu, Binh and Arnold, Guido and Richter, Marcus and Lippert, Th and Watanabe, Hiroshi and Ito, Nobuyasu},
  journal={Computer Physics Communications},
  volume={176},
  number={2},
  pages={121--136},
  year={2007},
  publisher={Elsevier}
}

@inproceedings{xu2024atlas,
  title={Atlas: Hierarchical partitioning for quantum circuit simulation on gpus},
  author={Xu, Mingkuan and Cao, Shiyi and Miao, Xupeng and Acar, Umut A and Jia, Zhihao},
  booktitle={SC24: International Conference for High Performance Computing, Networking, Storage and Analysis},
  pages={1--17},
  year={2024},
  organization={IEEE}
}

@inproceedings{haner2017,
  title={5 petabyte simulation of a 45-qubit quantum circuit},
  author={H{\"a}ner, Thomas and Steiger, Damian S},
  booktitle={Proceedings of the International Conference for High Performance Computing, Networking, Storage and Analysis},
  pages={1--10},
  year={2017}
}

@article{jones2019quest,
  title={QuEST and high performance simulation of quantum computers},
  author={Jones, Tyson and Brown, Anna and Bush, Ian and Benjamin, Simon C},
  journal={Scientific reports},
  volume={9},
  number={1},
  pages={10736},
  year={2019},
  publisher={Nature Publishing Group UK London}
}

@inproceedings{zhang2015quantum,
  title={Quantum computer simulation on multi-GPU incorporating data locality},
  author={Zhang, Pei and Yuan, Jiabin and Lu, Xiangwen},
  booktitle={International Conference on Algorithms and Architectures for Parallel Processing},
  pages={241--256},
  year={2015},
  organization={Springer}
}

@inproceedings{niwa2002general,
  title={General-purpose parallel simulator for quantum computing},
  author={Niwa, Jumpei and Matsumoto, Keiji and Imai, Hiroshi},
  booktitle={International Conference on Unconventional Methods of Computation},
  pages={230--251},
  year={2002},
  organization={Springer}
}

@article{fu2024achieving,
  title={Achieving energetic superiority through system-level quantum circuit simulation},
  author={Fu, Rong and Su, Zhongling and Zhong, Han-Sen and Zhao, Xiti and Zhang, Jianyang and Pan, Feng and Zhang, Pan and Zhao, Xianhe and Chen, Ming-Cheng and Lu, Chao-Yang and others},
  journal={arXiv preprint arXiv:2407.00769},
  year={2024}
}



\end{document}